(references added)

\documentstyle[aps,epsf,12pt]{revtex}
\newcommand{\be}{\begin{equation}}
\newcommand{\ee}{\end{equation}}
\newcommand{\bea}{\begin{eqnarray}}
\newcommand{\eea}{\end{eqnarray}}

\newcommand{\p}{\partial}

\tightenlines 

\begin{document}

\draft
\preprint{\small}

\title{Scalar field collapse in three-dimensional AdS spacetime}

\author{Viqar Husain$ ^\dagger$ and Michel Olivier\footnote{
Email addresses: husain@math.unb.ca, molivier@phy.ulaval.ca.}}

\address{\baselineskip=1.4em $ ^\dagger$Department of Mathematics and Statistics, \\
University of New Brunswick,
Fredericton NB E3B 1S5, Canada.\\
$ ^*$ D\'{e}partement de Physique,\\
Universit\'{e} Laval, 
Qu\'{e}bec, QC G1K 7P4, Canada. }

\maketitle 

\bigskip

\begin{abstract}
We describe results of a numerical calculation of circularly 
symmetric scalar field collapse in three spacetime dimensions 
with  negative cosmological constant. The procedure uses a double 
null formulation of the Einstein-scalar equations. We see evidence 
of  black hole formation on first implosion of a scalar pulse  
if the initial pulse amplitude $A$ is greater than a critical value 
$A_*$. Sufficiently near criticality the apparent horizon radius 
$r_{AH}$ grows with pulse amplitude according to the formula 
$r_{AH} \sim (A-A_*)^{0.81}$. 

\end{abstract}
\bigskip
 
\section{Introduction} 
Dynamical gravitational collapse of matter field configurations 
continues to be an interesting problem in general relativity.
The  questions concern whether the long time static or stationary 
limit of collapse results in a black hole or a naked singularity, 
the details of the final outcome, and the dependence of results 
on matter type and symmetries of the system. 
 
Among the simplest such systems is the Einstein-scalar equations 
with minimally coupled massless scalar field with certain 
symmetries. Initial work on this system in four spacetime dimensions 
in spherical symmetry was due to Christodoulou \cite{christo}, who 
showed that there exist classes of initial scalar configurations which 
give regular solutions for all time. Subsequently, numerical work by 
Goldwirth and Piran \cite{GP} established that black hole formation 
occurs for certain classes of initial data. A few years later a 
comprehensive study of this system by Choptuik \cite{MC} answered 
several important questions. In particular 
he showed that regular initial data characterized by parameter $p$ 
leads to a black hole with mass $M \sim (p-p_*)^\gamma$ where 
$p_*<p$ is a certain critical value of $p$ and $\gamma$ is a 
numerically determined exponent. This work also showed that the 
solution $p=p_*$ is a naked singularity. 

Since this initial work many other systems have been studied and 
similar behaviour found at the threshold of black hole formation. 
A semi-analytical understanding of the exponent in the mass formula 
has been obtained by assuming a critical solution and studying the 
linearized perturbation equations. For a comprehensive review and 
references see Ref. \cite{CG}. 

In the present work we describe a numerical simulation of minimally 
coupled scalar field collapse in three spacetime dimensions with 
negative cosmological constant. This system is of interest for a 
number of reasons. (i) A length 
scale set by the cosmological constant is present 
unlike the four dimensional Einstein-scalar case, where the only scale 
(apart from the gravitational constant) enters via initial data. 
(ii) There is a static black hole solution 
in three dimensions\cite{BTZ} with mass determined in units  
of the (negative) cosmological constant. In particular, there is no 
black hole with zero cosmological constant.(iii) The asymptotic 
spacetime is anti-deSitter (AdS) and not Minkowski, which may permit 
the possibility of reflecting an outgoing scalar field back to the 
interior.  

Furthermore, on  a separate but related note, there is currently 
interest in asymptotically anti-deSitter metrics from string theory, 
where the so called AdS/CFT correspondence has been proposed between 
gravitational phenomena in the bulk of anti-deSitter  spacetime and 
physical effects in certain conformally invariant Yang-Mills theories 
on its boundary \cite{JM}. The essence of the correspondence are certain 
proposed `dictionary' entries which translate between gravitational and 
Yang-Mills phenomena. In particular, the correspondence relates low energy 
gravity processes with high energy Yang-Mills ones, and vice versa. Although 
this correspondence is for five dimensional AdS, there is an analog of the 
correspondence relevant to three dimensions. Thus, any critical or other 
behaviour found in matter collapse in the bulk AdS spacetime may provide hints 
for a dictionary entry for the  AdS/CFT conjecture. As such results of the 
present work are potentially of interest for string theory.

There has been other work, both analytical and numerical on matter 
collapse in three dimensions. There are Vaidya-like solutions \cite{vh}
for collapsing dust, with and without pressure. Analytic results 
have been found for the collapse of dust shells \cite{PS} and 
point particles \cite{Mat,BS}. A recent numerical 
study of this same system has been carried out by Pretorius and 
Choptuik using a different coordinatization and numerical technique 
\cite{pc}. More recently an exact critical self-similar solution 
has been found \cite{Gcrit}, and the nature of the singularity 
within the black hole has been studied using a model system \cite{LB}.

This paper is organized as follows. The next section presents the 
system of equations and the details of the boundary conditions we use. 
Section III describes  the numerical procedure. The final section 
describes the results and outline of possible future work. 

\section{Einstein-Scalar equations}
The  Einstein-scalar equations with cosmological constant 
$\Lambda\equiv -1/l^2$ are 
 \be
  G_{ab} - {1\over l^2}\ g_{ab} = 2\pi T_{ab},
\ee
with 
\be 
T_{ab} = \p_a\phi \p_b\phi - {1\over 2} g_{ab} g^{cd}\p_c\phi \p_d\phi.
\ee 
These may be rewritten in the form
\be 
R_{ab} + {2\over l^2}\ g_{ab} = 2\pi \p_a\phi \p_b\phi. 
\ee

The general circularly symmetric  metric in three spacetime 
dimensions in double null coordinates $(u,v)$ may be written as 
\be
ds^2 = g_{ab}\ dx^adx^b = -2 l\ g(u,v)\ r'(u,v)\ dudv 
+ l^2\ r^2(u,v)\ d\theta^2,
\ee
where we are taking $g$ and $r$ to be dimensionless functions, $(u,v)$ 
and $l$ have dimensions of length, and 
$f' = \p f/\p v$, $\dot{f} = \p f/ \p u$.
In this parametrization, coordinate freedom has been fixed since there 
are only two metric functions to be determined.  Time evolution may 
be viewed as occuring with respect to the coordinate $u$. 

The $vv$ component of the Einstein equations is the constraint 
\be 
{g'r'\over gr} = 2\pi (\phi')^2 , 
\label{cons}
\ee
on a fixed $u$ surface. The $\theta\theta$ component is the time 
(or $u$) evolution of $r$
\be 
 l\dot{r}' = - rr'g.
\label{rdot0}
\ee
Finally, the time evolution of the scalar field is provided by 
the Klein-Gordon equation
\be 
(r\phi'\dot{)\ } + (r\dot{\phi})' = 0. 
\label{kg}
\ee
The remaining components of the Einstein equation are identically 
satisfied as consequence of these three equations. 

If $\phi=0$ these equations admit a black hole solution first 
discovered in \cite{BTZ}. In the standard $(r,t)$ coordinates 
the black hole metric is 
\be 
ds^2 = -\left(\left({ r\over l}\right)^2 - M\right)dt^2 +
         \left(\left({r\over l}\right)^2 - M\right)^{-1} dr^2 
         + r^2 d\theta^2. 
\label{met}
\ee
AdS space is recovered by setting $M=-1$. We can label the 
outgoing radial null geodesics of the AdS metric by $u$ and the 
ingoing ones by $v$. These coordinates are defined by 
\be 
u = t - l\ {\rm tan}^{-1}\left({r\over l}\right)  \ \ \ \ \ \ \ \ 
v = t + l\ {\rm tan}^{-1}\left({r\over l}\right)  
\ee
Therefore AdS space in double null coordinates with our
choice of dimensions and parametrization is given by 
\be 
r(u,v) = \ {\rm tan} \left({ v-u\over 2l}\right), \ \ \ \ \ \ \ 
g(u,v) =1.
\ee
Similarly, the BTZ black hole in the corresponding null coordinates 
is  given by \footnote{Note that this is the BTZ solution for 
$r < \sqrt{M}$. For $r> \sqrt{M}$ the tanh is replaced by coth.}
\be 
r(u,v) = \sqrt{M}\ {\rm tanh}\left[ {\sqrt{M} \over 2l}(u-v) \right],
\ \ \ \ \ \ \ g(u,v) = 1.
\label{btzuv}
\ee
Thus $r=0$ corresponds to $u=v$. In the following 
we will use this as one of the boundary conditions for the full 
set of coupled equations.  
 
Equations (\ref{cons}--\ref{kg}) may be rewritten in a first order 
initial value form suitable for numerical integration as follows:
On a constant $u$ surface, integrating the constraint 
equation (\ref{cons}) gives 
\be 
g =  k_1(u)\ {\rm exp}\left[ 2\pi \int_u^v d\tilde{v}
\left({r\over r'} (\phi')^2\right) \right],
\label{g}
\ee
where $k_1(u)$ is an arbitrary function of $u$.

Integrating (\ref{rdot0}) gives 
\be 
\dot{r} = - {1\over l}\int_u^v d\tilde{v}\ (grr') + { k_2(u)\over l},
\label{rdot}
\ee
where $k_2(u)$ is an arbitrary function of $u$. Note that in 
both of these equations we are integrating from $u$ to $v$ 
which is to correspond to integrating outward from $r=0$. 

Finally,  using $\dot{r}$ from (\ref{rdot}), and the 
definition
\be
 h\equiv {2r\phi' + r'\phi\over r'},
\label{h}
\ee
the Klein-Gordon equation (\ref{kg}) gives
\be 
\dot{h} = {r\over l}\ (h-\phi) \left( g  + {k_2(u) 
- \bar{g}\over 2r^2}\right),
\label{hdot}
\ee
where 
\be 
\bar{g}(u,v) \equiv  
\int_u^v d\tilde{v}\ \left(g
r r' \right).
\ee

The definition of $h$ Eqn. (\ref{h}) gives
\be 
\phi(u,v) = {1\over 2\sqrt{r}} \int_u^v d\tilde{v}\  
\left( {h\ r'\over \sqrt{r}} \right)  + {k_3(u) \over \sqrt{r} }.
\label{phi}
\ee
The function $k_3(u)$ is fixed by noting from Eqn. (\ref{h}) 
if $\phi = C$ then $h=C$, where $C$ is a constant. Comparing 
this with Eqn. (\ref{phi}) gives $k_3(u) = 0$.
  
The equations (\ref{g}), (\ref{rdot}), and (\ref{hdot}) give 
the required null initial value formulation for the three-dimensional 
Einstein-scalar system with negative cosmological 
constant $\Lambda = -1/l^2$. These equations contain two arbitrary 
functions $k_1(u)$ and $k_2(u)$, which are to be fixed by boundary 
conditions. With these functions fixed, initial data sets   
$r(u=0,v)$ and $h(u=0,v)$ (derived from $\phi$) can be evolved 
to later times $u$. 

Let us now see how the arbitrary functions in the equations 
are fixed. The most natural boundary condition is the requirement that the 
metric is regular at the origin $r=0$, which is a timelike line. In 
our formulation $r(u,v)$ is a field variable. Therefore we must first  
define where $r=0$ is in order to impose regularity conditions there. 
As noted above the natural choice motivated by AdS and the BTZ black hole 
is  $r=0$ at $u=v$, i.e. 
\be 
r(u,u)=0,
\ee
which implies 
\be 
\dot{r}(u,u) = -r'(u,u).
\ee
With this choice for $r=0$, there are two possible boundary conditions 
that may be imposed at $r=0$. Before describing 
these we set $k_1(u) = 1$ for simplicity.

\noindent {\bf (i)} No conical singularity at $r=0$. This gives the 
condition
  \be
      \dot{r}(u,u) = -n^2g(u,u)/ 2l.   
  \label{bc1}
  \ee
where $n$ is an integer. (See the Appendix.)
Using this in Eqn. (\ref{rdot}) gives 
\be
k_2(u) =  - {n^2 g(u,u)\over 2} = -{n^2\over 2},
\label{bc11}
\ee
which is to be used in the evolution equations (\ref{rdot}) and 
(\ref{hdot}).
 
\noindent {\bf (ii)} The $uv$ metric function is unity: 
\be
    2l\ g(u,u)\ r'(u,u) = - 2 l g(u,u)\ \dot{r}(u,u) = 1.
    \label{bc2}
\ee
Therefore 
\be \dot{r}(u,u) = -{1\over 2 lg(u,u)},
\ee
and eqn. (\ref{rdot})  gives 
\be 
k_2(u) = -{ 1\over 2}
\label{bc22}
\ee
We shall use $k_2(u) = -1/2$. In summary the first order equations 
to be integrated  numerically are 
\bea
\dot{r} &=& -{\bar{g}\over l} - {1\over 2l} \label{rdotf}\\
\dot{h} &=& {r\over l}\ (h-\phi) \left( g - {\bar{g}\over 2r^2} 
- {1\over 4r^2} \right) 
\label{hdotf}
\eea
where 
\bea
g &=& {\rm exp}\left[ {\pi\over 2} \int_u^v d\tilde{v}
\left({r'\over r} (h - \phi)^2\right) \right] \\
\bar{g} &=& \int_u^v d\tilde{v}\ \left(gr r' \right) \\
\phi &=& {1\over 2\sqrt{r}} \int_u^v d\tilde{v}\  
\left( {h\ r'\over \sqrt{r}} \right).
\eea

Notice that with $h =\phi=0$ our equations with $k_2$ 
an arbitrary constant reduce to     
\be 
\dot{r}(u,v) = -{r^2 - 2k_2\over 2l},
\ee
which is to be integrated with the initial condition $r(u,u)=0$.
The solution is 
\be 
r(u,v) = \sqrt{2k_2}\ {\rm tanh}\left[{\sqrt{2k_2}\over 2l}(u-v)\right].
\ee
Comparing with (\ref{btzuv}), we see that 
\be 
M = 2k_2.
\ee
Thus the value $k_2 = -1/2$, which we use for numerical integration, 
corresponds to  $M=-1$, which is 3-dimensional AdS. Notice that although 
this is the natural case to study in that it is most similar to the 
spherically symmetric collapse in four dimensions (because $r=0$ is 
a regular point initially), the other choices for $k_2$ corresponding 
to $M=0$, $M>0$ and $ -1 < M < 0$ are also interesting.  

\section{Numerical Method}

Our numerical scheme is based on the approach used in \cite{GP},  
and its refinements in \cite{DG}. 

The evolution equations (\ref{rdotf}--\ref{hdotf}) are converted 
into set of coupled ordinary differential equations on a grid by the 
correspondence 
\be 
h(u,v) \rightarrow h_n(u),\ \ \ \ \ \ \ r(u,v)\rightarrow r_n(u),
\ee
and similarly for the derived variables $g(u,v)$, $\bar{g}(u,v)$, 
and $\phi(u,v)$ where $n = 0,\cdots, N$
specifies the $v$ grid. The differential equations are therefore 
 \bea
\dot{r}_n &=& -{\bar{g}_n\over l} - {1\over 2l} \label{rdotd}\\
\dot{h}_n &=& {r_n\over l}\ (h_n-\phi_n) 
\left( g_n - {\bar{g}_n\over 2r_n^2} - {1\over 4r_n^2} \right) 
\label{hdotd}
\eea

We take the initial value functions to be  
\be 
r(0,v) = v,  
\ee 
with $h(0,v)$ determined by this and the initial scalar field 
configuration
\be 
\phi(u=0,r) = A r^2\ {\rm exp}\left(-{r-r_0\over \sigma^2}\right), 
\ee
where $A,\sigma$ and $r_0$ are initial data parameters of the 
scalar field. The initial specification of the other functions is 
completed by computing the integrals for $g_n$ and $\bar{g}_n$ using 
Simpson's rule.

The boundary conditions at fixed $u$ are 
\be 
r_k=0, \ \ \ \bar{g}_k=0,\ \ \ \ g_k=1.  
\ee
{\it where $k$ is the index corresponding to the position of the origin
$r=0$.} (In the algorithm used all grid points $ 0\le i \le k-1$ 
correspond to ingoing rays that have reached the origin and 
are dropped from the grid; see below). These conditions are equivalent 
to $r(u,u)=0$, $g|_{r=0} = g(u,u) =1$ etc., and guarantee regularity of 
the metric at $r=0$.  Notice that for our initial data, $\phi_k$ and 
hence $h_k$ are initially zero, and therefore remain zero at the origin 
because of Eqn. (\ref{hdotd}). 

Our numerical procedure is similar to that used in Refs. \cite{GP,DG}.
Time evolution with respect to the coordinate $u$ is initiated via a 
fourth order Runge-Kutta scheme. The constraint integrals are recalculated 
on each $u$ slice. As evolution proceeds the entries in the $r_n$ array 
sequentially reach $r=0$, at which point they are dropped from the grid. 
Thus the radial grid loses points with evolution. This is similar to the 
procedure used in \cite{GP} and \cite{DG}. At each $u$ step, a check is made 
to see if an apparent horizon has formed. This is done in the standard way 
by calculating and plotting the function 
\be 
g^{ab}\p_ar\p_br = -{\dot{r}\over lg}.
\ee
Evolution terminates if either an apparent horizon forms, or the 
scalar field pulse reflects off $r=0$. In practice we stop the 
evolution if $|\p_u r|$ becomes $10^{-4}$. 

To  improve numerical accuracy near $r=0$  
we follow a procedure similar to \cite{DG}, where all functions 
on a constant $u$ surface are expanded in power series in $r$ at 
$r=0$, and the first three values of the constraint integrals are 
calculated using the respective power series. We write 
\be 
h = h_0 + h_1 r
\ee
from which the following relations are derived:
\be 
\phi = h_0 + {h_1r\over 3}, \ \ \ \ \ g = 1+{\pi h_1^2r^2\over 9}, 
\ \ \ \ \  \bar{g} = {r^2\over 2} + {r^2 {\rm ln}g\over 4}.
\ee
In practice, because of our boundary conditions, $h_0$ turns out 
to be of order $10^{-5}$. The remaining $N-3$ values of these 
functions are computed using Simpson's rule for equally spaced 
points. This substantially improves accuracy near $r=0$. 
$v-$derivatives of functions are calculated using 
$f'_i = (f_{i+1} - f_{i-1})/{2\Delta v}$ with end point values 
determined by linear extrapolation: $f'_1 = 2f'_2 - f'_3$ and 
$f'_{N} = 2f'_{N-1} - f'_{N-2}$. 
 
All numerical calculations were performed for $l=1$ with up to 1400
v-grid points and scalar field parameters 
$\sigma = 0.3$ and $r_0 = 1.0$.  The code is written in C and 
was executed on a Sun workstation. 

Before describing our results in the next section, we comment 
further on the numerical procedure outlined above.  

Firstly,  because we use double null coordinates with the number of 
$v$ grid points decreasing as ingoing null geodesics 
cross $r=0$, we are not able to follow a reflected pulse back out 
toward infinity.  This is likely a shortcoming because after first implosion
some percentage of the scalar field intensity will be reflected
back toward infinity and reach it in finite time (due to the compactness
of AdS space). It is then possible that some
part of the field would be reflected back toward the origin from the
boundary at infinity. This would tend to increase the mass of the
black hole.  Multiple reflections could occur several times until 
a static limit is reached. Our procedure is limited to evaluating the 
mass of the black hole on first implosion of the scalar field
pulse. We can assume however that the mass increase due
to multiple
reflections would be small compared to the mass obtained on first
implosion. 

Secondly, again due to the loss of points from the grid, we are not able to
observe the behaviour of the scalar field very near criticality. However, 
the calculations performed in
\cite{pc} indicate that there is continuous 
self similarity
in the near critical regime.
\footnote{ As in our case, and 
for similar reasons, this work also focusses on a search for critical and 
scaling behaviour on first implosion of the scalar pulse.}
 
\section{Results and Comments}
 
After several simulations for different values of the initial amplitude $A$
of the scalar field, we arrive at the following conclusions. For amplitudes
smaller than $A_* \simeq 0.1248$ the field is reflected back in totality
from the origin and no black hole is formed. For $A > A_*$ however, we observe
black hole formation. 
\begin{figure}[t]
\epsfxsize=140mm
\hspace*{5mm}
\epsffile{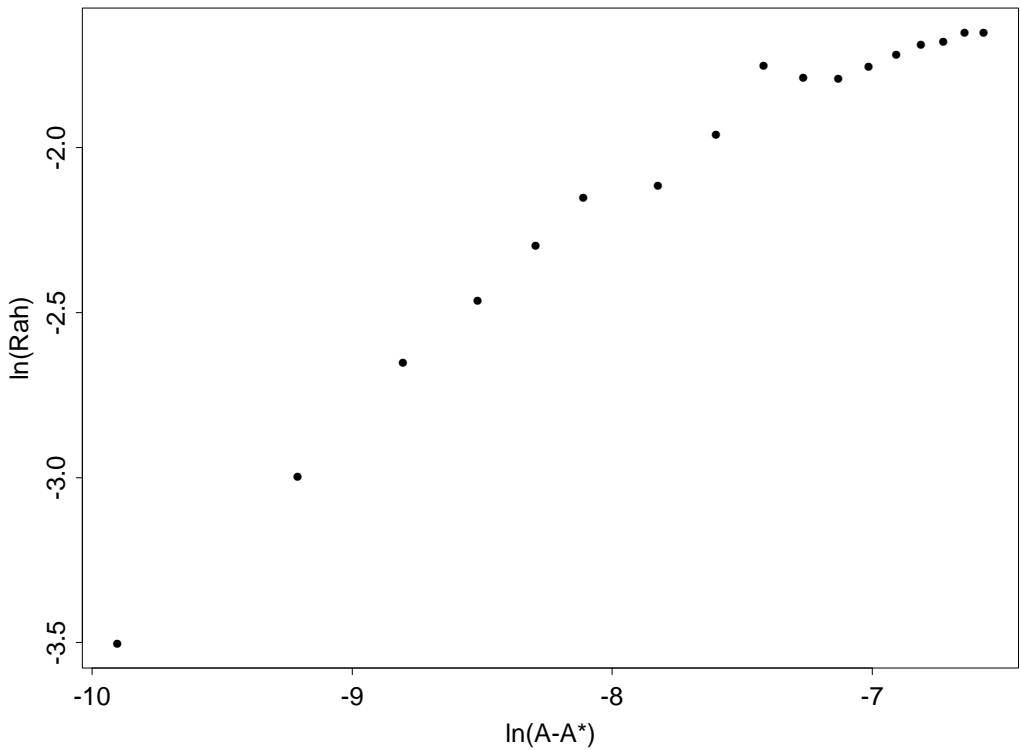}
\vspace*{1mm}
\caption{Logarithmic plot of the apparent horizon radius $r_{AH}$ versus
the initial amplitude of the imploding scalar field $(A-A_*)$.} 
\end{figure}
Figure 1
shows a logarithmic plot of the
apparent horizon radius $r_{AH}$ versus the initial amplitude $A$. 
The plot shows that for $A$ sufficiently close to $A_*$ the behaviour is nearly
linear (the first six points). This is the region of interest for us since it
appears to exhibit the type of scaling behaviour observed in many other collapse
studies \cite{MC,CG}. We  ignore the other part of the plot which shows some
oscillatory behaviour and eventually a flattening of the curve, because this 
is further from criticality. We thus estimate
the value of the critical exponent by considering the linear part. This yields
the result

\begin{equation}
r_{AH} \sim (A-A_*)^\gamma
\end{equation}

\noindent where $\gamma \simeq 0.81$. Given that black-hole mass is
proportional to $r_{AH}^2$ for the BTZ black-hole, our mass exponent estimate
is $1.62$. Pretorius and Choptuik \cite{pc} used a different method proposed
by Garfinkle to estimate the critical exponent. They studied the subcritical
($A<A_*$) scaling behaviour of the maximum value of the scalar curvature $R$ to
obtain a value for $\gamma$ in the range $1.15-1.25$.
  
There are several directions for future study using the double null 
procedure outlined here. These include computations for other 
values of the constant $k_2$ and other initial scalar configurations.
In addition, the self similar behaviour of the scalar field near criticality 
found numerically in \cite{pc} and analytically in \cite{Gcrit}, can likely 
be reproduced using the method of grid point restoration used in \cite{DG}.

In connection with the AdS/CFT conjecture, it will be useful to carry 
out similar calculation for 5-dimensional AdS space. The computational 
procedure used here requires relatively minor modifications to 
study that case.

\bigskip
\noindent{\it Acknowledgements:} We thank Matt Choptuik, Frans Pretorius, and 
Bill Unruh for many helpful comments and discussions. This work was supported 
by the Natural Science and Engineering Research Council of Canada (NSERC)
and the Fonds pour la Formation de Chercheurs et l'Aide \`{a} la
Recherche (FCAR).

\section{Appendix}
The condition of no conical singularity at the origin $r=0$ is derived. 
The parallel transport of a vector $A^a$ on a closed loop around the origin
should be such that this vector comes back to itself after a complete turn. 
The condition for parallel transport of $A^a$ along a curve 
tangential to $t^a$ is
\be 
t^a \nabla_a A^b = 0.
\ee 
Moving on  a curve of constant $u$ and $v$, only $\theta$ varies
from $0$ to $2\pi$, i.e. $t^a = (\partial/\partial\theta)^a$ and the above
equation becomes

\begin{equation}
\label{partrans}
\nabla_\theta A^b = \partial_\theta A^b + \Gamma^b_{\theta c} A^c = 0 .
\end{equation}
The relevant  non-zero $\Gamma$  are

\begin{eqnarray}
\Gamma^u_{\theta\theta} = \frac{r}{g} \qquad \Gamma^v_{\theta\theta} = \frac{r\dot{r}}{gr'} \\
\Gamma^\theta_{\theta u} = \frac{\dot{r}}{r} \qquad \Gamma^\theta_{\theta v} = \frac{r'}{r} .
\end{eqnarray}

\noindent The parallel transport equation is then 
 
\begin{eqnarray}
\begin{array}{cc@{\:=\:}cc}
\partial_\theta & \left[ \begin{array}{c} A^u \\ A^v \\ A^\theta \end{array}
\right] &
\left[ \begin{array}{ccc} 0&0&-\frac{r}{g} \\ 0&0&-\frac{r\dot{r}}{gr'}\\
-\frac{\dot{r}}{r}&-\frac{r'}{r}&0 \end{array} \right] & \left[ \begin{array}{c}
A^u \\ A^v \\ A^\theta \end{array} \right] \end{array} .
\end{eqnarray}

\noindent Diagonalizing  the matrix on the right we get 3 eigenvalues $\lambda = 0,
\pm \sqrt{\frac{2\dot{r}}{g}}$. In the eigenvectors basis we would thus have
the following differential equation to solve for each component

\begin{equation}
\partial_\theta A^\alpha = \lambda^{(\alpha)} A^\alpha .
\end{equation}

\noindent This is solved by $A^\alpha (\theta) = A^\alpha (0) \exp
\left( \lambda^{(\alpha)} \theta \right)$. Requiring that $A^\alpha$ be unchanged
 after a complete non-trivial turn around the origin requires 

\begin{equation}
A^\alpha(2\pi) = A^\alpha(0) \Rightarrow A^\alpha(0)\exp\left(2\pi\lambda^{(\alpha)}\right)
= A^\alpha(0) \Rightarrow \lambda^{(\alpha)} = ni ,
\end{equation}

\noindent where $n \in$ Z. The simplest non-trivial case is
$\lambda^{(\alpha)}=i$, and this becomes a boundary condition if we look at the
case where $\alpha$ is the third eigenvalue :

\begin{equation}
\sqrt{\frac{2\dot{r}}{g}} = i \qquad  \mbox{at} \qquad r=0 .
\end{equation}

\noindent This first boundary condition is thus

\begin{equation}
\dot{r} = -\frac{g}{2} \qquad \mbox{at} \qquad r=0 .
\end{equation}

\end{document}